\documentclass[twocolumn,prb]{revtex4}

\usepackage{graphicx}
\usepackage{bm}
\usepackage{epstopdf}
\usepackage{amsmath,amsfonts,paralist}

\let\a=\alpha \let\b=\beta  \let\g=\gamma  \let\d=\delta \let\e=\varepsilon
     \let\l=\lambda
                
\let\s=\sigma    \let\f=\varphi 
   
\let\G=\Gamma \let\D=\Delta  \let\L=\Lambda

\def\to{\rightarrow}

\newcommand{\beq}{\begin{equation}}
\newcommand{\eeq}{\end{equation}}

\newcommand{\Tr}{\text{Tr}}

\begin{document}

\title{On the two-steps relaxation of mean-field glasses: $p$-spin model}

%\title{Parisi - Picco - Ritort model at criticality: threshold values of $M$
%and $p$ between continuos and discontinuous transition} 

\author{U. Ferrari$^{1,2}$, L. Leuzzi$^{1,2}$,
G. Parisi$^{1,2,3}$ and T. Rizzo$^{1,2}$} \affiliation{$^1$ Dip. Fisica,
Universit\`a {\em La Sapienza}, Piazzale A. Moro 2, I-00185, Rome, Italy \\
$^2$ IPCF-CNR, UOS Rome {\em Kerberos}, Universit\`a {\em La Sapienza}, Piazzale A. Moro 2,
I-00185, Rome, Italy \\ $^3$ INFN, Piazzale A. Moro 2, 00185, Rome,
Italy}

\begin{abstract}
Critical slowing down dynamics of supercooled glass-forming liquids is usually understood at the mean-field level in the framework of  Mode Coupling Theory, providing a two-time relaxation scenario and   power-law behaviors of the time correlation function at dynamic criticality. 
In this work we derive critical slowing down exponents of spin-glass models undergoing discontinuous transitions by computing their Gibbs free energy and connecting the dynamic behavior to static ``in-state'' properties. Both the spherical and Ising versions are considered and, in the simpler spherical case, a generalization to arbitrary schematic Mode Coupling kernels is presented.
Comparison with   dynamic results  available in literature is performed. Analytical predictions for the  Ising case are provided for any $p$.
\end{abstract} 
\date{\today} \maketitle

\section{Introduction}

The slowing down of the dynamics of supercooled glass-forming liquids  corresponds to a non trivial underlying thermodynamic landscape. 
An unusual time behavior of the density correlation function emerges, with respect to the exponential decay: a separation between fast ($\b$) and 
slow ($\a$) relaxation modes  takes place and the correlation function develops a plateau approaching a dynamic arrest transition. \cite{Leuzzi07b,Cavagna09,Goetze09}

Theoretical advances  have been made studying such systems with the so called Mode Coupling Theory (MCT), \cite{Bengtzelius84,Goetze84,Goetze89, Biroli04, Goetze09,Bouchaud96} a mean-field theoretical description of many particle systems, 
able to identify the separation of two relaxation processes and a \textit{dynamical transition temperature} $T_d$ at which ergodicity breaks down with the system undergoing a structural arrest.
In this framework, for $T \to T_d$ the time spent at the plateau increases and diverges at $T_d$. At criticality a  power-law behavior arises for the correlation close to the plateau and the exponents governing the approach to ($\sim t^{-a}$) and the depart from ($\sim t^b$) the plateau are related as:
\beq
\frac{\Gamma^2(1-a)}{\Gamma(1-2 a)}=\frac{\Gamma^2(1+b)}{\Gamma(1+2 b)}=\lambda ~,
\label{eq_MCT}
\eeq
where $\l$ is a model dependent quantity (functional of the static structure factor \cite{Weysser10}), that is usually  treated like a tunable parameter.

Since the works of Kirkpatrick, Thirumalai and Wolynes, \cite{Kirkpatrick87a,Kirkpatrick87b,Kirkpatrick87c,
Kirkpatrick88a,Thirumalai88,Kirkpatrick89} the behavior of glass-forming liquids and structural glasses has been linked with dynamics and thermodynamics of a certain class of mean-field spin-glass (SG) models, sometimes called \textit{mean-field} glasses or  \textit{discontinuous} spin-glasses. 
These  include  the $p$-spin models, \cite{Gardner85,Crisanti92,Crisanti93} displaying a dynamic transition at which ergodicity breaks down, with a behavior of the correlation function close or identical to the one predicted by MCT. 
Below this dynamic transition the Boltzmann measure splits in many well defined metastable states, which at a temperature $T_s$ become thermodynamically stable inducing a phase transition. \cite{Kirkpatrick87b,Crisanti92,Cavagna05}
The static transition  is governed by an entropy crisis and plays the role of the hypothesized Kauzmann transition, related to the vanishing of the configurational entropy of the liquid. \cite{Cavagna98,Gibbs58, Leuzzi07b}
Below this point, named Random First Order Transition (RFOT), the model develops a spin-glass phase with one step  \textit{replica symmetry breaking} (1RSB).
Besides the conjectured thermodynamic  analogy between discontinuous spin-glasses and structural glasses, cf.  \textit{Mosaic Theory}, \cite{Kirkpatrick89, Lubchenko07, Berthier11} the former can as well  be exploited   for studying critical slowing down dynamics and MCT properties at the dynamic arrest transition.

In this paper we do not address to the study of the thermodynamic but we investigate only the dynamical transition: we apply to the $p$-spin model a recent development in the comprehension of the dynamical transition and its relation to the thermodynamic behavior, \cite{Caltagirone11} i.e., 
a relationship between critical slowing down exponents and thermodynamic ``in-state" quantities: \cite{Caltagirone11, Rizzo12}
\begin{equation}
 \l = \frac{w_2}{w_1} ~,
 \label{f:lambda_w}
\end{equation}
with the $\l$ defined in Eq. (\ref{eq_MCT}). 

The coefficients $w_1$, $w_2$ are obtained from the expansion at the third order terms of the appropriate Gibbs free-energy $\Gamma[q]$  functional of the  order parameters, i.e. the replica overlap matrix elements $q_{\a\b}$. The potential $\Gamma$ is the Legendre  transform the  free energy 
functional as function of replica "pinning" fields.\cite{Monasson95,Franz98b, Franz11b} It coincides with the Franz-Parisi potential,\cite{Franz95a} by means of which the dynamic transition is identified as the spinodal point of an excited local minimum at a non-zero overlap value equal to the plateau value of the correlation function. 
Critical slowing down exponents can, then, be obtained from the coefficients of the expansion of the $\Gamma$ potential around a replica symmetric (RS) solution $q^{RS}$ at the plateau value of the correlation function at the dynamic critical point.

For $T \gtrsim T_d$ the Gibbs potential can be expanded in powers of overlap fluctuations around the dynamic solution, $\delta q_{\a\b}\equiv Q_{\a\b}-Q_{\a\b}^{RS}$:
\begin{eqnarray}
 \d \G[ \d q_{\a\b}] &\sim& \sum_{(\a \b),(\g \d)} \! \! \! M_{\a \b,\g \d} ~ \d q_{\a \b} \d q_{\g \d} + \nonumber \\
  && \sum_{(\a \b) (\g \d) (\e \f)} \! \! \! \! \! W_{\a \b \g \d \e \f} ~\d q_{\a \b} \d q_{ \g \d} \d q_{ \e \f}~,
 \end{eqnarray}
where $\d \G[\d q_{\a\b}] = \G[Q_{\a\b}]  -  \G[Q_{\a\b}^{RS}] $ and the first order term is absent at the saddle-point. Here and below brackets in sums over replica indices mean that only distinct replicas must be considered. 

The dynamic transition is associated to the vanishing of one eigenvalue of the mass term, the \textit{replicon}, which defines a critical direction in the replica space. 
Projecting this expansion along the replicon direction $\delta^{(r)}q_{ab}$ (see next session for the details) one obtains:
\begin{equation}
\d \G[ \d^{(r)} q_{\a\b}] \sim w_1 \Tr (\d^{(r)} q)^3 + w_2 \sum_{(\a \b)} (\d^{(r)} q_{\a \b})^3 ~.
\label{G_3}
\end{equation}
%where $\d \G[\d^{(r)} q_{\a\b}]  =\G[q^* + \d^{(r)} q_{\a\b}] - \G[q^*]$ [{\bf che differenza  tra $q^*$ e $q^{RS}$?}].  

In order to obtain the MCT exponents for mean-field discontinuous SG models one can proceed with the following  protocol:
\begin{enumerate}
 \item compute the averaged replicated action $\G[Q]$;
 \item compute the expansion around the dynamical RS solution up to the third order, cf. Eq. (\ref{G_3});
 \item solve the system equation for the saddle point  and the vanishing of the replicon, both in the $n \to 1$ limit which allows to work on the dynamic metastable state of the Gibbs potential;  
 \item evaluate the coefficients $w_1$, $w_2$ of the third order along the replicon direction;
 \item compute $\l$ by means of Eq. (\ref{f:lambda_w}) and $a$ and $b$ with Eq. (\ref{eq_MCT}).
\end{enumerate}

\section{The spherical p-spin model}
In this section we focus on the spherical version of the fully-connected $p$-spin model. Its relevance in the context of mean-field glasses is due to the fact that for temperatures down to the dynamic transition,
the dynamic equations for its two time correlation functions  \cite{Kirkpatrick87c, Crisanti93} are equivalent to those of the  schematic MCT. \cite{Goetze89, Bouchaud96} The Hamiltonian of the model reads:
\beq
\mathcal{H} = - \sum_{i_1 < \dots < i_p} J_{i_1 \dots i_p} \sigma_{i_1} \dots \sigma_{i_p} - h \sum_i \sigma_i~,
\label{f:Ham_psph}
\eeq
where the couplings $J$ are Gaussian independent identically distributed variables with:
\beq
P(J)=\sqrt{\frac{N^{p-1}}{\pi  p!}}  \exp \left( -\frac{N^{p-1}}{  p!} J^2 \right) \label{distr_J}
\eeq
and the spins $\sigma$ are real values variables  subject to a global constraint:
\beq
\sum_i \sigma_i^2 = N ~.
\eeq
Due to the spherical constraint this model is analytically solvable in all details needed for our scope. This allow us to show how to compute analytically third order coefficients and dynamic exponents $a$ and $b$ step by step. 

Through a saddle point calculation, it is possible to compute the replicated  partition function: \cite{Crisanti92}
\begin{eqnarray}
\overline{Z^n} &=& e^{S(\infty)} \int_{|q|>0} \prod_{\alpha < \beta} \sqrt{\frac{N}{2 \pi}} dq_{\alpha \beta} \exp \left\{ - N \G[Q] \right\} \nonumber
\\
 \G[Q] &=&  -\frac{\mu}{2p} \sum_{\alpha \beta} Q_{\alpha \beta}^p - \frac{(\b h)^2}{2} \sum_{\a \b} Q_{\a\b}  \nonumber \\
 &&- \frac{1}{2} \log |Q| + \frac{(\b h)^4}{2} \left(  \sum_{\a \b} Q_{ab} \right)^2 ~ ,
\label{G[q]}
\end{eqnarray}
where the overbar means average over disorder, cf. Eq. (\ref{distr_J}), $\mu = p \beta^2/2$, $Q_{\alpha \beta} = 1/N \sum_i \sigma_i^\alpha \sigma_i^\beta$ and 
\beq
e^{S(\infty)} =e^{N(1+\log(2 \pi))/2} \pi^{-1/2} \left[1 + O\left(\frac{1}{N} \right)\right]~.
\eeq
Differentiating Eq. (\ref{G[q]}) with respect to $Q_{\a \b}$ one gets the saddle point condition:
\begin{equation}
 \mu Q_{\a \b}^{p-1} + (\b h)^2 + (Q^{-1})_{\a \b} =0 ~, \quad \forall \a \neq \b
 \label{f:sp_spher}
\end{equation}
where the term proportional to $h^4$ is absent because it is irrilevant in both the $n\to 0,1$ limits.

For our purpose we also need the  third order Taylor expansion of $\G$ with respect to fluctuations around the RS solution: $\d q_{\a \b} = Q_{\a \b} - Q^{RS}_{\a \b}$, with $Q^{RS}_{\a \b}= (1-q)\d_\a^\b+q$,where $\d_\a^\b$ is the Kronecker delta:
\begin{eqnarray}
 2\d \G[\d q]& \simeq &  \frac{1}{2!} \sum_{(\a \b)( \g \d)}\!\! \G''_{\a \b \g \d} \d q_{\a \b} \d q_{ \g \d}
 \nonumber \\
&+& \frac{1}{3!}\sum_{(\a \b) (\g \d) (\e \f)}\!\!\! \G'''_{\a \b \g \d \e \f} \d q_{\a \b} \d q_{ \g \d} \d q_{ \e \f}~,%+O(\d q^4)
\label{deltaG}
\end{eqnarray}
where $\d \G[\d q]= \G[Q]- \G[Q^{RS}]$. The first order is absent at the saddle point and
\begin{eqnarray}
 &&\G''_{\a \b \g \d} \equiv \frac{\partial^2 \Gamma}{\partial Q_{\a\b}\partial Q_{\g \d}}=
\\
 \nonumber 
 &&\qquad - (p-1) \mu Q_{\a \b}^{p-2} \d_\a^\g \d_\b ^\d 
    +(Q^{-1})_{\a \g} (Q^{-1})_{\d \b} + (\b h)^4\\
&& \G'''_{\a \b \g \d \e \f} =\frac{\partial^3 \Gamma}{\partial Q_{\a\b}\partial Q_{\g \d}\partial Q_{\e\f}} =
\label{G3}
\\
\nonumber
&&\qquad- (p-1)(p-2) \mu Q_{\a \b}^{p-3} \d_\a^\g \d_\b ^\d \d_\a^\e \d_\b ^\f 
\\
\nonumber
&&\qquad - 2 (Q^{-1})_{\a \e} (Q^{-1})_{\f \g} (Q^{-1})_{\d \b}~.
\end{eqnarray}
In order to study the critical dynamic behavior we  work within a RS Ansatz with $n \to 1$  \cite{Monasson95,Franz98b,Crisanti08,Franz11b} and restrict our analysis to the replicon subspace, defined by the conditions
\begin{equation}
 \sum_{\a} \d q_{\a \b} = \sum_{\b} \d q_{\a \b}=0~,
\label{replicon_subsp}
\end{equation}
(the second condition is a consequence of the first one as far as $q_{\a \b}$ is a symmetric matrix. The vanishing of the eigenvalue of the Hessian in this subspace, the so-called \textit{replicon} eigenvalue, yields the criticality condition.

We now restrict the analysis to the case without external magnetic field $h=0$.
Imposing the saddle point condition, cf. Eq. \ref{f:sp_spher}, at the dynamic transition point (RS with $n \to 1$)
\begin{equation}
 \mu q^{p-1} = \frac{q}{(1-q)} \label{sp_sph}
\end{equation}
and  the vanishing of the replicon
\begin{equation}
(p-1)\mu q^{p-2} = \frac{1}{(1-q)^2} \label{repl_sph}~,
\end{equation}
leads to the expressions for  the value of the overlap at the dynamic transition and the dynamic temperature:
\begin{equation}
q_d = \frac{p-2}{p-1} ~, \quad \mu_d = \frac{(p-1)^{p-1} }{(p-2)^{p-2} }~.
\label{sol_dyn}
\end{equation}
The dynamic transition point is also the point at which the dynamic saddle point solution appears, that is,
Eqs. (\ref{sp_sph})-(\ref{repl_sph}) are not independent  and, indeed, Eq. (\ref{repl_sph}) can be obtained also as the derivative of the Eq. (\ref{sp_sph}).

Considering fluctuation only in the replicon subspace, one can considerably simplify the expansion of Eq. (\ref{deltaG}) as
\begin{eqnarray}
 2\d \G[\d q^{(r)}] &\simeq& - (p-1)(p-2) \mu q^{p-3}\sum_{(\a \b)} (\d q^{(r)}_{\a \b})^3\nonumber \\
 && - \frac{2}{(1-q)^3} \sum_{(\a \b \g)} \d q^{(r)}_{\a \b} \d q^{(r)}_{\b \g} \d q^{(r)}_{\g \a}~,
\end{eqnarray}
In this case the tensorial form of Eq. (\ref{G3}) is so simple that we can straightforwardly compute the values of the cumulants $w_1$ and $w_2$, yielding
\begin{equation}
 \frac{ w_2 }{ w_1 } = \frac{ (p-1)(p-2) }{2} \mu q^{p-3} (1-q)^3  ~,
 \label{w2w1_sph}
\end{equation}
which, imposing Eq. (\ref{repl_sph}), reduces to:
\begin{equation}
 \frac{ w_2 }{ w_1 } = \frac{(p-2)(1-q)}{2 q} ~ .
\label{lambda_simply}
\end{equation}
Using the value of $q_d$ at the transition, cf Eq. (\ref{sol_dyn}), one obtains for $\lambda$ the $p$-independent value
\begin{equation}
 \l = \left. \frac{ w_2 }{ w_1 }\right|_d = \frac{1}{2} ~,
\end{equation}
which coincides with the result reported in Ref. [\onlinecite{Crisanti93,Franz11}].

\subsection{The case of uniform magnetic field}
The presence of a magnetic field term ($h \neq 0$) can change the nature of the transition. 
As shown in Refs. [\onlinecite{Crisanti92,Crisanti93}] for values of the field $h >h_{tr}= \sqrt{p^{2-p} (p-2)^p/2}$ the transition becomes continuos:  no plateau is there and
at the transition the long time limit of the correlation function does not jump discontinuously, though  the relaxation behavior in time is still a power-law in the $\b$ regime. The exponent $a$ (sometimes called $\nu$) is the only one defined.
% and, in this case, is usually called $\nu$. 
We first study this continuous transition and then we move to the discontinuous one.

The value of $a$ can be computed as for the discontinuous transition case, except for the fact that now one has to work in the $n\to 0$ limit.\cite{Caltagirone11,Rizzo12} 
This implies that the RS expression of Eq. (\ref{f:sp_spher}) becomes
\begin{equation}
 \mu q^{p-1} + (\b h)^2 = \frac{q}{(1-q)^2} ~.
\end{equation}
rather than Eq. (\ref{sp_sph}).
Equations (\ref{repl_sph}) and (\ref{w2w1_sph}), instead, do not change. From Ref. [\onlinecite{Crisanti92}] we know that the transition line is parametrically defined as
\begin{eqnarray}
T^2 &=& \frac{p(p-1)}{2} (1-q)^2 q^{p-2} \label{trans_param_T}\\
h^2 &=&  \frac{p (p-2)}{2} q ^{p-1}
\label{trans_param_h}
\end{eqnarray}
with $1-2/p < q < 1$. Given a value of the field $h$, one can, thus, straightforwardly compute the corresponding values of $q$ and $T$ from Eq. (\ref{trans_param_T}) and (\ref{trans_param_h}) and obtain $\l$ from 
Eq. (\ref{lambda_simply}).

For the discontinuous transition in a field $h<h_{tr}$, 
 two non-zero overlap  values are relevant:  the plateau value $q_1$ and the long-time limit $q_0=C(\infty)$.  
Comparing with the results of Ref.   [\onlinecite{Crisanti92}] we observe that the exponent parameter $\l$ is still given by the ratio $w_2/w_1$, cf.  Eq. (\ref{lambda_simply}), now evaluated on $q=q_1$, 
solution of the following equations for the dynamic critical values of $T_d(h)$, $q_0(h)$ and $q_1(h)$: 
\begin{equation}
\frac{1}{(1-q_1)^2}=\mu(p-1)q_1^{p-2}, ~~\mbox{ cf. Eq.  (\ref{repl_sph})}
\nonumber
\end{equation}
and
\begin{eqnarray}
\frac{1}{1-q_1}-\frac{1}{1-q_0}&=&\mu\left(q_1^{p-1}-q_0^{p-1}\right)
\nonumber
\\
\mu q_0^{p-1}&=&\frac{q_0}{(1-q_0)^2}-(\b h)^2  \nonumber ~,
\end{eqnarray}
which are the 1RSB saddle point equations (cf. Eq.  (\ref{f:sp_spher}))

\subsection{Generalization to arbitrary schematic MCT models}
Comparing the dynamics of the $p$-spin spherical model \cite{Crisanti93} with the MCT differential equation for the correlation function, \cite{Goetze89,Bouchaud96} one can notice that the function $\mu q^{p-1}$ ($\mu \phi^{p-1}$, in the MCT notation), derivative of the first term of the action (\ref{G[q]}), plays the role of the MCT memory kernel. \cite{Bouchaud96,Kirkpatrick87b,Crisanti93} One can generalize this argument,through schematic MCT, to a generic polynomial kernel\cite{Goetze09}
\begin{equation}
\L(\phi(t)) = \mathcal{F}[\{v\},\phi(t)] = \sum_p v_p \phi(t)^{p-1}.
\label{f:Lambda_kernel}
\end{equation}
We start considering the long-time limit of the correlation function, $q_d$  (else called non-ergodicity parameter $f$), and the $\phi(t)$ expansion around such limit: 
\begin{equation}
\phi(t) \simeq q_d + G(t); \qquad |G(t)|<< 1~,
\end{equation}
whose
  Laplace transform reads 
\begin{eqnarray}
\phi(z) &\sim& -\frac{q_d}{z} + G(z) \label{phi_lapl} ~.
\end{eqnarray}
 These behaviors have to satisfy the MCT dynamical equation and its transformed:
\begin{eqnarray}
\tau \partial_t \phi(t) + \phi(t) &=& - \int_0^t du \L(t-u) \partial_u \phi(u) \\
\frac{\phi(z)}{1+z \phi(z)} &=& i \tau + \L(z) \label{MCTeq_lapl}~.
\end{eqnarray}
Plugging Eq. (\ref{phi_lapl}) into the left hand side of Eq. (\ref{MCTeq_lapl})  and expanding up to the third order, one obtains
\begin{eqnarray}
&z \frac{\phi(z)}{1+z \phi(z)} \sim& \nonumber \\
& \frac{ - q_d + z G(z)}{1- q_d + z G(z)} \left[1 - \frac{z G(z)}{1- q_d} + \frac{z^2 G(z)^2}{(1- q_d)^2}+ O(G^3) \right] &\label{left}~.
\end{eqnarray}
The Laplace transform of the expansion of the  memory kernel reads:
\begin{eqnarray}
z \L(z) &\sim& - F(v, q_d) + F'(v, q_d)G(z) \nonumber \\
&& +  \frac{1}{2} F''(v, q_d) \mathcal{LT}[ G(t)^2](z) \label{right}~.
\end{eqnarray}
One has to equate equations (\ref{left}) and (\ref{right}) order by order in $G(t)$. The zeroth and  first order are the standard long-time MCT equation and its derivative, cf. Eqs. (\ref{sp_sph}),(\ref{repl_sph}):
\begin{eqnarray}
F(v, q_d) &=& \frac{q_d}{1-q_d}  \label{f:mct_exp1}\\
F'(v, q_d) &=& \frac{1}{(1-q_d)^2} \label{f:mct_exp2} ~.
\end{eqnarray}
The third order yields:
\begin{equation}
z G(z)^2 + \l z \mathcal{LT}[ G(t)^2](z) =0
\end{equation}
with:
\begin{equation}
\l \equiv \frac{1}{2} F''(v, q_d) (1- q_d)^3~.
\label{f:lambdaMCT}
\end{equation}
Assuming a power-law solution $G(t) \sim  \left( t_0/t \right) ^{a,-b}$ one gets back Eq. (\ref{eq_MCT}) with a parameter exponent coinciding with Eq. 
(\ref{f:lambdaMCT}).

This result can also be obtained  studying a model whose action is a slight generalization of Eq. (\ref{G[q]}) action:
\begin{eqnarray}
\G[Q] &=&  -\frac{1}{2 } \sum_{\alpha \beta}A( Q_{\alpha \beta}) -\frac{1}{2} \log |Q| 
\label{genericAction}
\end{eqnarray}
such that $\Lambda(x)=A'(x)$.
For polynomial function $A(x)$ this action describe a model with an Hamiltonian composed by a proper sum of $p$-spin interaction terms, like Eq. 
(\ref{f:Ham_psph}). \cite{Crisanti04b,Crisanti07b, Crisanti11}
From the action Eq. (\ref{genericAction}) one can easily derive Eqs. (\ref{f:mct_exp1}), (\ref{f:mct_exp2}), with a kernel given by Eq. (\ref{f:Lambda_kernel}),
yielding critical temperature (identical  to the mode coupling temperature) and critical plateau value of the correlation.
%\begin{eqnarray}
 %\L_T(q) &=& \frac{q}{(1-q)} \\
%\L'_T(q) &=& \frac{1}{(1-q)^2} \label{LAMBDA_repl_sph}
%\end{eqnarray}
Expanding Eq. (\ref{genericAction}) to third order, cf. Eq. (\ref{deltaG}), yields:
\begin{equation}
\l = \frac{w_2}{w_1}=\frac{1}{2} \L''(q_d) (1-q_d)^3
\end{equation}
coinciding with Eq. (\ref{f:lambdaMCT})
and verifying the method proposed for the schematic MCT models. This is the simplest case, where equations of motion can be solved and analytic results are available. We now move to consider a more difficult case for which $\lambda$ cannot be computed directly solving the dynamic equations 
and Eq. (\ref{f:lambda_w}) remains the only way, known so far, to estimate the critical slowing down exponents.

\section{The Ising p-spin model}
\label{sec:ising}
In this section we focus on the Ising version of the fully-connected $p$-spin model. The Hamiltonian of the model reads:
\beq
\mathcal{H} = - \sum_{i_1 < \dots < i_p} J_{i_1 \dots i_p} \sigma_{i_1} \dots \sigma_{i_p}~,
\eeq
where the couplings $J$ are Gaussian distributed again with Eq. (\ref{distr_J}) and  $\sigma_i= \pm 1$. It is well known \cite{Gardner85} that this model displays first a RFOT with a dynamical transition $T_d$ separated from a static transition $T_s$ to a spin-glass 1RSB stable phase and, at a lower temperature,  a second transition to a spin-glass full RSB phase. For our purpose, we only focus  on the first transition.

Averaging over disorder and introducing the overlap matrix through an auxiliary matrix $\L_{ab}$, one gets the replicated action:
\begin{eqnarray}
 \G[Q,\L] &=& \frac{ \beta^2}{4} \sum_{(\a \b)} Q_{\a \b}^p  -\frac{1}{2}\sum_{(\a \b)} \L_{\a \b} Q_{\a \b} \nonumber \\
 &&+\log \Tr_{\{\sigma\}} ~ \mathcal{W}[\L;\sigma] \label{ising_action}\\ 
\mathcal{W}[\L;\sigma] &=& \exp \left(  \frac{1}{2} \sum_{(\a \b)} \Lambda_{\a \b} \sigma_\a \sigma_\b  \right)
\end{eqnarray}
that has to be evaluated through a saddle point calculation. The derivative with respect to  $\L_{\a\b}$ yields
\begin{equation}
\frac{p \b^2}{2} Q^{p-1}_{\a \b} = \L_{\a \b}, \quad \a \neq \b
\label{lambda_saddle}
\end{equation}
leading to
\begin{eqnarray}
 \G[Q] &=& - \frac{(p-1)\beta^2}{4} \sum_{(\a \b)} Q_{\a \b}^p  +\log \Tr_{\sigma}  ~\mathcal{W}[Q;\sigma]\nonumber  \\ 
\mathcal{W}[Q;\sigma] &=&  \exp \left(  \frac{p \b^2}{4} \sum_{(\a \b)} Q^{p-1}_{\a \b} \sigma_\a \sigma_\b \right) \label{actionQ}~.
\label{action_is}
\end{eqnarray}
This concludes the first step of our protocol. 
The first order derivative reads:
\begin{equation}
\G'_{\a \b} = \frac{\partial \G}{\partial Q_{\a\b}}=\theta Q_{\a \b}^{p-2}\left(  <\sigma_\a \sigma_\b> - Q_{\a \b}  \right)~,
\label{first_ord}
\end{equation}
where $\theta= p(p-1) \b^2/4$ and $< \dots >$ means average over the weight $\mathcal{W}[Q]$, cf. Eq. (\ref{actionQ}). The vanishing of this equation gives the saddle point condition, that, solved in the RS Ansatz with $n \to 1$, yields
\begin{eqnarray}
q &=& < \hat{m}^2>=\mathcal{N}^{-1}\int dz ~\mathcal{W}[z] \tanh(z)^2 \label{q_saddle_ising}\\
\nonumber
\hat m &=& \tanh(z)
\\
\mathcal{W}[z] &=& \exp \left(- \frac{z^2}{p \b^2 q^{p-1}} \right)  \cosh(z)\\
\mathcal{N} &=& \!\!\! \int \! dz~ \mathcal{W}[z] = \sqrt{\pi p} \b q^\frac{p-1}{2} \exp \left( \frac{p \b^2 q^{p-1}}{4} \right) \nonumber
\label{sp_ising}
\end{eqnarray}

The second order derivative reads:
\begin{eqnarray}
\G''_{\a \b, \g \d} &=&\!\!\! \theta (p-2) Q_{\a \b}^{p-3} \d_\a^\g \d_\b^\d \left(  <\sigma_\a \sigma_\b> - Q_{\a \b}  \right) 
 \label{second_ord}
\\
 &+&\!\!\! \theta Q_{\a \b}^{p-2} \left( \theta Q_{\a \b}^{p-2}  <\sigma_\a \sigma_\b \sigma_\g \sigma_\d>_c -  \d_\a^\g \d_\b^\d   \right)~,
\nonumber 
\end{eqnarray}
where the presence of connected averages $\langle \ldots \rangle_c$ is a consequence of the direct derivative of the term $\log \Tr_{\sigma}  ~\mathcal{W}[Q;\sigma]$ (cf. App. \ref{appcumulants}).
In order to impose criticality, $\G''_{\a \b, \g \d}$, evaluated at the saddle point condition and projected onto the replicon subspace, should vanish.  
The first part of Eq. (\ref{second_ord}) is proportional to $\G'_{\a\b}$, cf. Eq. (\ref{first_ord}) and it does not contribute.  The vanishing of the second part, in the RS Ansatz, reads
%\begin{equation}
% \theta Q_{\a \b}^{p-2}  (<\sigma_\a \sigma_\b \sigma_\g \sigma_\d>_c^R)^{sp} -  \d_\a^\g \d_\b^\d=0
%\label{replicon}
%\end{equation}
%which, as explained in appendix \ref{appcumulants}, in the RS Ansatz is equivalent to:
\begin{equation}
2 \theta q^{p-2}  (1-2q +r) -  1=0 ~,
\label{RSreplicon}
\end{equation}
as detailed in App. \ref{appcumulants}. Here $r= < \hat{m}^4>$.
This allows to rewrite Eq. (\ref{RSreplicon}) as:
\begin{eqnarray}
\frac{1}{2} &=& \theta q^{p-2}  <(1- \hat{m}^2 )^2> = \nonumber \\
&=& \theta q^{p-2}  < \mbox{sech}^4 (x) > ~.
\end{eqnarray}
Once the saddle point and the vanishing of the second order derivative (Eq. (\ref{second_ord})) are imposed, the third order derivative reads:

\begin{eqnarray}
\frac{Q_{\a \b}^{6-3p}}{ \theta^3}\G'''_{\a \b, \g \d, \e \f} &=& \frac{(p-2) Q_{\a \b}^{3 - 2p}}{\theta^2} \d_\a^\g \d_\b^\d \d_\a^\e \d_\b^\f \nonumber \\ 
&& + <\sigma_\a \sigma_\b \sigma_\g \sigma_\d \sigma_\e \sigma_\f>_c
\end{eqnarray}
and this allows to write the coefficients of the expansion, cf. Eq. (\ref{G_3}), as
\begin{eqnarray}
\frac{q_d^{6-3p}}{ 8 \theta^3}w_1&=&  1- 3 q_d + 3 r_d -u_d =\nonumber \\
&=& < ( 1-\hat{m}^2)^3 > \label{om1_is}\\
\frac{q_d^{6-3p}}{ 8 \theta^3}w_2 &=& 2( q_d - 2 r_d + u_d) + \D = \nonumber \\
 &=&  2 < \hat{m}^2 (1-\hat{m}^2)^2 > +  \D  \label{om2_is} \\
 \D &=&  \frac{2 (p-2) q_d^{3-2p}}{\b_d^4 p^2 (p-1)^2} \label{delta_om2} ~,
\end{eqnarray}
where $u= <\hat{m}^6>$. As it happens in the Sherrington-Kirkpatrick (SK) model \cite{Sompolinsky82,Caltagirone11}, the term $\D$ vanishes if $p=2$ and, indeed, can be considered as the correction to $w_2$ due to the multi-body interaction.  

With this results one can obtain the numerical values of the coefficients and the exponent $a$ reported in Tab. I.
%  \begin{table}
% \label{tab:Isip}
%  \begin{tabular}{c|c|cccccccccc|c} 
%  $p$ & $\to 2$ & $2.05$ &$2.2$&$2.5$&$3$&$4$&$5$&$6$&$7$&$8$&$9$&$\to \infty$  \\
%  \hline
%  $\lambda$ &$.5$&$.556$&$.652$&$.719$&$.743$&$.746$& $.743$&$.739$& $.736$&$.733$&$.731$&$.666$ \\
% \hline 
% $a$ &$.395$& $.379$&$.346$&$.32$&$.308$&$.306$&$.308$&$.31$&$.311$&$.313$&$.314$&$.340$\\
% \hline
%  $T_d$ &$1$&$.916$&$.808$&$.724$&$.682$&$.678$& $.700$&$.727$& $.756$&$.784$&$.812$&$\infty$ \\
% \hline 
%  $q_d$ &$0$&$.051$&$.198$&$.428$&$.643$&$.815$& $.881$&$.915$& $.935$&$.948$&$.957$&$1$ \\
% \hline 
%  \end{tabular}
%  \caption{Dynamic exponents in the Ising $p$-spin model.}
%   \vskip -.3cm \end{table}
% \begin{table}
%\label{tab:Isip}
% \begin{tabular}{c|c|cccccccc|c} 
 %$p$ & $\to 2$ & $2.05$ &$2.2$&$2.5$&$3$&$4$&$5$&$6$&$9$&$\to \infty$  \\
 %\hline
 %$\lambda$ &$.5$&$.556$&$.652$&$.719$&$.743$&$.746$& $.743$&$.739$&$.731$&$.666$ \\
%\hline 
%$a$ &$.395$& $.379$&$.346$&$.32$&$.308$&$.306$&$.308$&$.31$&$.314$&$.340$\\
%\hline
%$b$ &$\slash$& $.892$&$.768$&$.609$&$.57$&$.565$&$.57$&$.576$&$.589$&$.67$\\
%\hline
 %$T_d$ &$1$&$.916$&$.808$&$.724$&$.682$&$.678$& $.700$&$.727$&$.812$&$\sqrt{p/(4 \log p) }$ \\
%\hline 
 %$q_d$ &$0$&$.051$&$.198$&$.428$&$.643$&$.815$& $.881$&$.915$&$.957$&$1$ \\
%\hline 
% \end{tabular}
 %\caption{Dynamic exponents in the Ising $p$-spin model.}
  %\vskip -.3cm \end{table}
 \begin{table}
\label{tab:Isip}
 \begin{tabular}{c|ccccc} 
 $p$&$T_d$&$q_d$&$\l$&$a$&$b$ \\
 \hline
  $\to 2$&$1$&$0$&$1/2$&$.395$&$1$\\
 \hline
   $2.05$&$.916$&$.051$&$.556$&$.379$&$.892$\\
 \hline
   $2.2$&$.808$&$.198$&$.652$&$.346$&$.768$\\
 \hline
   $2.5$&$.724$&$.428$&$.719$&$.320$&$.609$\\
 \hline
   $3$&$.682$&$.643$&$.743$&$.308$&$.570$\\
 \hline
   $4$&$.678$&$.815$&$.746$&$.307$&$.565$ \\
 \hline
   $5$&$.700$&$.881$&$.743$&$.308$&$.570$ \\
 \hline
   $6$&$.727$&$.915$&$.739$&$.310$&$.576$ \\
 \hline
   $7$&$.756$&$.935$&$.736$&$.311$&$.581$ \\
 \hline
   $8$&$.784$&$.948$&$.733$&$.313$&$.586$ \\
 \hline
   $9$&$~~.812~~$&$~~.957~~$&$~~.731~~$&$~~.314~~$&$~~.589~~$ \\
 \hline  
   $\to \infty$&$\sqrt{\frac{p}{4 \log p}}$&$1$&$2/3$&$.340$&$.700$ \\
   \hline 
 \end{tabular}
 \caption{Dynamic exponents in the Ising $p$-spin model.}
  \vskip -.3cm \end{table}
  
%\begin{equation}
% \begin{array}{c|c|c}
% p & \frac{w_2}{w_1} & a \\ 
%\hline
%2.05 & 0.556 & 0.379 \\ 
%2.2 & 0.652 & 0.346 \\
%2.5 & 0.719 & 0.32 \\
%3 & 0.743 & 0.308  \\
%4 & 0.746 & 0.306 \\
%5 & 0.743 & 0.308 \\
%6 & 0.739 & 0.310 \\
%7 & 0.736 & 0.311 \\
%8 & 0.733 & 0.313 \\
%9 & 0.731 & 0.314
%\end{array} \nonumber
%\end{equation}

\subsection{The $p \to 2$ limit}
The interest to the behavior of the model for $p$ close to two  is due to its relation with the SK model.
Doing an expansion for small $\epsilon = p - 2$, one can expect and, actually, self-consistently verify  
that the finite jump of the overlap at the transition is of order $\epsilon$. One can consequently expand the action for small $Q_{\a \b}$, still considering the transition as discontinuous.\\
From Eq. (\ref{lambda_saddle}) one obtains
\begin{equation}
\frac{1}{\b^2} \L^{1-\e}_{\a \b} = Q_{\a \b} ~.
\label{lambda_saddle_eps}
\end{equation}
Putting this result in the action (\ref{ising_action}) and expanding for small $\L_{\a \b}$ one gets:
\begin{eqnarray}
 \G[\L] &\sim& - \frac{1}{4\beta^2} \sum_{\a \b} \L_{\a \b}^{2-\e}  +\frac{1}{4}\sum_{\a \b} \L^2_{ \a \b} \nonumber \\
 && +\frac{1}{6} \Tr \L^3 + O(\L^4) ~.
\end{eqnarray}
The first three derivatives read:
\begin{eqnarray}
2\G'_{\a \b} &=& - \frac{2-\e}{2 \b^2} \L^{1-\e}_{\a \b} + \L_{\a \b} + (\L^2)_{\a \b}\\
 2\G''_{\a \b, \g \d} &=& - \frac{(2-\e)(1-\e)}{2 \b^2} \L^{-\e}_{\a \b} \d_\a^\g \d_\b^\d  \nonumber \\
  && + ~ \d_\a^\g \d_\b^\d  + \d_\a^\g \L_{\d \b} + \L_{\a \g} \d_\b^\d \\
2\G'''_{\a \b, \g \d, \e \f} &=& \frac{(2-\e)(1-\e) \e}{2 \b^2} \L^{-1-\e}_{\a \b} \d_\a^\g \d_\b^\d \d_\a^\e \d_\b^\f \nonumber \\
 && + ~ \d_\a^\g \d_\d^\e \d_\b^\f + \d_\a^\e \d_\g^\f \d_\b^\d \quad \label{G3_eps}.
\end{eqnarray}
From the first two derivatives one gets the criticality condition. In the RS Ansatz and with $n\to 1$ this reduces to the system equation:
\begin{eqnarray}
 \frac{1}{\b^2} \hat\l^{1-\e} &=& \hat\l -  \hat\l^2 \\
 \frac{1- \e}{\b^2} \hat\l^{-\e} &=&  1 - 2 \hat\l ~,
\end{eqnarray}
where $\hat \l$ is the off-diagonal part of $\L_{\a \b}$. The system equation is solved by:
\begin{eqnarray}
\hat\l_d&\sim& \e \\
\b_d &\sim& 1- \frac{1}{2}\e \log \e ~ .
\end{eqnarray}
Evaluating the third order derivative (\ref{G3_eps}) on this result one gets:
\begin{equation}
 \l = \left. \frac{w_2}{w_1} \right|_d= \frac{1}{2} ~ 	.
\end{equation}
This result agrees with the $a=.395$ proposed by Kirkpatrick and Thirumalai\cite{Kirkpatrick87c} studying the dynamics of a soft-spin version of the model in the $p \to 2$ limit.

We note that the behaviour of the parameter $\lambda$ is discontinuous as a function of $p$ at $p=2$. Indeed in the SK model, {\it i.e.} precisely at $p=2$, we have $\lambda=w_2=0$, while as soon as $\epsilon>0$ we have $\lambda=1/2$. This happens because the coefficient $w_2$ is proportional to the third derivative of $q^{2+\epsilon}$ which is singular at $q=0$ as soon as $\epsilon$ is different from zero.

% The discontinuity with the SK model ($p=2$), where $\l = w_2=0$, is due to the term $\D w_2$ of Eq. (\ref{om2_is}), which has a finite $\e \to 0$ limit. Furthermore the presence of a discontinuity is not surprising: 
% unlike the continuous $p=2$ case, where the system equation must be treated with $n\to0$, for any finite $\e$ the transition is discontinuous and the same equations must be analyzed in the $n\to1$ limit.   

\subsection{The $p \to \infty$ limit}
The behavior of the $p$-spin model for large $p$ has been previously studied, due to its relation to the Random Energy Model (REM).\cite{Derrida80,Derrida81,Gross84} 
In this section we, indeed, present the calculation of $\l= w_2/w_1$ in the limit $p \to \infty$.

As in the $p\to 2$ case, here is more convenient to work with the auxiliary variable $\hat \l$, the RS off-diagonal element of the matrix $\L_{\a \b}$, which is related to $q$ by Eq. (\ref{lambda_saddle}). 
Furthermore in order to keep finite q, one should expect and consistently verifies that $\hat \l$ diverges in the large $p$ limit.  From Eq. (\ref{q_saddle_ising}) one gets:
\begin{eqnarray}
q(\hat \l)&=& \left( \frac{ 2 \hat \l}{p \b}\right)^\frac{1}{p-1} = < 1- (1-\hat m^2) >  \nonumber \\
&\simeq&1 - e^{ - \l/2 } \sqrt{2 \pi \l}~, \quad \hat \l \gg 1
\end{eqnarray}
where, in the last expression, only the leading term, for $\hat \l \gg 1$, has been retained. Differentiating this equation, one arrives to the system equation for the criticality condition. At the leading order it reads:
\begin{eqnarray}
q(\hat \l)&=&1 - e^{ - \hat\l/2 } \sqrt{2 \pi \hat\l} \label{q_p_inf}\\
\frac{2 q(\hat \l)}{p \hat\l} &=& e^{ - \hat\l/2 } \sqrt{2 \pi \hat\l} 
\end{eqnarray}  
Solving this system first for $\hat \l$ and then for $q$, which contains the whole dependence from the temperature, one finds:
\begin{eqnarray}
\hat \l_d &\simeq& 2 \log p\\
T_d &\simeq& \sqrt{ \frac{p}{4 \log p} }~.
\end{eqnarray}
As $p\to \infty$ the critical dynamic temperature diverges and the dynamic overlap $q_d \to 1$, cf. Eq. (\ref{q_p_inf}).

Now we move to the computation of the exponent parameter. Eq. (\ref{om1_is}), (\ref{om2_is}) and their ratio $\l$, in the large $\hat \l$ limit, reduce to:
\begin{eqnarray}
w_1 &=& \frac{3 \pi}{8}e^{ - \hat\l/2 } \sqrt{\frac{1}{2 \pi \hat\l}} \\
w_2 &=&\frac{\pi}{4}  e^{ - \hat\l/2 }\sqrt{\frac{1}{2 \pi \hat\l}} + \D  \\
\l &=& \frac{w_2}{w_1} = \frac{2}{3} \left( 1 + \D e^{ \hat\l/2 } \sqrt{\frac{32 \hat\l}{\pi} }   \right)
\end{eqnarray}
Evaluating the last term one obtains:
\begin{equation}
\D e^{ \hat\l/2 } \sqrt{\frac{32 \hat\l}{\pi} } \sim \sqrt{\frac{1}{2 \pi \log^3p}} \left( 1- \frac{1}{p} \sqrt{\frac{\pi}{4 \log p}}\right)^{-2p}
\end{equation}
which goes to zero as $p\to \infty$. 

Our result is, indeed, $\l \to 2/3$, corresponding to an exponent $a =.340$

\subsection{The addition of ferromagnetic couplings}
It is possible to generalize the previous results allowing the couplings $J$ to have a non zero mean:
\begin{equation}
 P(J) = \sqrt{\frac{N^{p-1}}{\pi p!}} \exp \left[ -\frac{N^{p-1}}{ p!} \left( J - \frac{p! J_0}{N^{p-1}}\right)^2 \right]~,
\end{equation}
%the parameter $J_0$ tunes the coupling average as:
%\begin{equation}
 %\ol{J} = \frac{p!}{N^{p-1}} J_0 \,.
%\end{equation}
To treat the case $J_0 \neq 0$ one has to introduce a non-zero magnetization for the system and the the Gibbs effective action, cf. Eq (\ref{action_is}), becomes:
\beq \begin{split}
 \G[Q,\Lambda, m, x] &= \frac{\beta^2}{4} \sum_{ab} Q_{ab}^p + \beta J_0 \sum_{a} m_{a}^p + \\
 &-\frac{1}{2}\sum_{ab} Q_{ab} \Lambda_{ab} - \sum_a m_a x_a   \\ 
&+ \log \Tr_{\{\sigma\}} ~\mathcal{W}[\L,x;\sigma]\\
\mathcal{W}[\L,x;\sigma]&\equiv
 \exp \left( \frac{1}{2} \sum_{ab} \Lambda_{ab} \sigma_a \sigma_b  +  \sum_a x_a \sigma_a \right) ~,
\end{split} \label{action_J0}\eeq
where the fields $x_a$ play for the magnetization, $m_a$,  the same role of $\L$ for the overlap. Through a saddle point calculation one arrives to two coupled equations:
\begin{eqnarray}
q &=& \langle \hat{m}^2\rangle= \mathcal{N}^{-1}\int dz ~\mathcal{W}[z] \tanh(z)^2\\
m &=& \langle \hat{m}\rangle =\mathcal{N}^{-1} \int dz ~\mathcal{W}[z] \tanh(z)\\
\mathcal{W}[z] &=& \exp \left(- \frac{(z -p \b J_0 m^{p-1})^2}{p \b^2 q^{p-1}} \right) \\
\mathcal{N} &=& \int dz~\mathcal{W}[z] = \sqrt{\pi p} \b q^\frac{p-1}{2}~.
%\label{sp_ising}
\end{eqnarray}
%where now the averages $\langle \cdot \rangle$ are computed with the weighting: $\exp{-\frac{(x- p \b J_0 m^{p-1})^2}{  p \b^2 q^{p-1}  } } / \sqrt{p \pi \b^2 q^{p-1} }$ and again $\hat{m}\equiv \tanh(x)$.  
The solution with $m=0$, the SG phase, is always present but, for $J_0$ large enough, a ferromagnetic (FM) RS solution appears discontinuously with $m=q>0$ and turns out to be the stable one. 

Since in the action (\ref{action_J0}) $J_0$ couples only to the magnetization $m$, the SG phase, where $m=0$, is not affected by the presence of a non-zero mean of the couplings. 
As a result, along the whole dynamic SG transition the physics does not change and the values of the exponents $a$ and $b$ are constant.

%The calculation of the main section is indeed correct up to this FM transition. 
The PM/FM transition is a usual thermodynamic first order transition with a ferromagnetic spinodal line. \cite{Nishimori01}
Two relevant point are, indeed, present: the tricritical point between the SG, the FM and the PM phases and the intersection between the dynamic transition line and the FM spinodal line. 
It has been shown \cite{Nishimori01,Krzakala11} that both the relevant points belong to the Nishimori Line (NL)
\beq
  J_0^{NL} (T) = \frac{1}{2 T}~,
\eeq
a line in the $J_0, T$ phase diagram. This fact is due to the following property of systems on the NL: \cite{Krzakala11}
\beq
\lim_{n \to 0} \left. \G(\b,J_0^{NL} (\b),q,m) \right|_{m=q} = \lim_{n\to 1} \G(\b,0,q,0) \, .
\label{NLidentity}
\eeq
Equation (\ref{NLidentity}) corresponds to the statement that the static Gibbs free energy potential along the NL is equal to the  dynamic Gibbs free energy along the $J_0=0$ axis. 
This means that, in order to obtain the ratio $w_2/w_1$ at the dynamical transition with $n \to 1$, one can, simplifying the numerical calculation, evaluate those coefficients along the NL, at the spinodal point and working with $n \to 0$. 
%, but still following Section \ref{sec:ising}. 
Furthermore, since on the NL the melting process is equivalent to a glassy transition, \cite{Nishimori01} one can argue that the exponent calculated through our procedure controls this melting process too.

\section{conclusions}
We have applied a new method \cite{Caltagirone11} to compute the slowing down exponents  on the Ising and spherical versions of the frustrated $p$-spin model with Gaussian interaction. 
This method allows us to derive the mode coupling exponents of the critical dynamics by means of an analytical static-driven computation. 
Those exponents govern the power-law approach to ($C(t) \sim q_d + t ^{-a}$) and departure from ($C(t) \sim q_d- t^b$) the plateau value $q_d$.

For the spherical case we exactly reproduce the analytical result obtained in Ref. [\onlinecite{Crisanti93}], derived by a Langevin description of the dynamics
and, in full generality, all schematic MCT exponents.

For the Ising version we present our computation and the exact values of both exponents for any value of $p$. In this case, the dynamics
with discrete spins has not been directly solved so far and ours are the first analytic estimates for $a$ and $b$.
One can, anyway, approximate discrete by soft spins in order to construct a dynamical equation. In that case the final computation for $p>2$ differs: the discrete case depends on the value of $p$, whereas the soft one does not.
Our results agree with the computation performed in Ref. [\onlinecite{Kirkpatrick87c}] on a soft-spin approximation in the  $p\to 2$ limit. In this limit we found a discontinuity with the Sherrington-Kirkpatrick model ($p=2$), with a finite jump, from $0$ to $1/2$ of the exponent parameter $\l$. For the sake of completeness, also the $p\to \infty$ limit has been characterized. The study of the Ising $p$-spin model concludes considering a non-zero mean of the couplings ($J_0\neq0$): up to a critical value, where a ferromagnetic transition takes place, the behavior of the model does not change with $J_0$, validating our estimates for the exponents along the whole SG transition line.

\acknowledgements
We thank F. Caltagirone, A. Crisanti, S. Franz , F. Ricci-Tersenghi and E. Zaccarelli for useful discussions.

\appendix
\section{Computation of the cumulants in the replicon sub-space}
\label{appcumulants}
In this appendix we present the calculation of the connected cumulants of four and six replicas projected in the replicon sub-space. 
At first should be noticed that in the RS Ansatz four replica index quantities could take only three different values, depending on how many replica indeces are repeated. For:
\begin{eqnarray}
&  C^{(4)}_{(\a \b) (\g \d)} = < \s_\a \s_\b \s_\g \s_\d >_c  = & \nonumber \\
   &= < \s_\a \s_\b \s_\g \s_\d > - < \s_\a \s_\b> <\s_\g \s_\d >&~,
\end{eqnarray}
we have:
\begin{eqnarray}
 C^{(4)}_{(\a \b) (\a \b)} &=& 1 - q^2\\
 C^{(4)}_{(\a \b)( \a \d)} &=& q - q^2 \\
 C^{(4)}_{(\a \b)( \g \d)} &=& r-q^2 ~,
\end{eqnarray}
 where here $\a,\b,\g,\d$ are considered different. This allows us to express the tensorial form of $C^{(4)}$:
\begin{eqnarray}
 C^{(4)}_{(\a \b)( \g \d)} &=& (1- 2 q +r) (\d_\a^\g \d_\b^\d +\d_\a^\d \d_\b^\g)  \nonumber \\
&&  + (q-r) (\d_\a^\g +\d_b^\d + \d_\a^\d +\d_b^\g )\nonumber \\
&& + (r-q^2) ~.
\end{eqnarray}
Summing over all replica indeces the $C^{(4)}$ times the fluctuations in the replicon sub-space ( Eq. (\ref{replicon_subsp}) ), the only term that does not vanish is $(1- 2 q+r)$, with a factor $2$ due to the exchange of $\a$ with $\b$:
\begin{eqnarray}
&&\sum_{(\a \b),(\g \d)} \!\! \!\! C^{(4)}_{(\a \b) (\g \d)} \d q^{(r)}_{\a \b} \d q^{(r)}_{\g \d} = \nonumber \\
&&=2(1- 2 q +r) \sum_{(\a \b)}  (\d q^{(r)}_{\a \b} )^2
\end{eqnarray}

Six replica index quantities, like:
\begin{eqnarray}
  C^{(6)}_{(\a \b) (\g \d)(\e \f)} &=& < \s_\a \s_\b \s_\g \s_\d \s_\e \s_\f>_c ~ =  \\
  &=& < \s_\a \s_\b \s_\g \s_\d \s_\e \s_\f> \nonumber \\
   &-& < \s_\a \s_\b \s_\g \s_\d > <\s_\e \s_\f>\nonumber \\
  &- & < \s_\g \s_\d \s_\e \s_\f>< \s_\a \s_\b> \nonumber\\ 
  &-&   < \s_\a \s_\b \s_\e \s_\f> < \s_\g \s_\d> \nonumber \\
  &+&  2 < \s_\a \s_\b > < \s_\g \s_\d > <\s_\e \s_\f >\nonumber  ~,
\end{eqnarray}
 can take eight different values. The computation follows as in the previous case \cite{Pimentel02} leading to:
\begin{eqnarray}
\sum_{(\a \b),(\g \d)(\e \f)}\!\!\!\!\!\! && \!\! \!\! C^{(6)}_{(\a \b) (\g \d)(\e \f)} \d q^{(r)}_{\a \b} \d q^{(r)}_{\g \d}\d q^{(r)}_{\e \f} = \nonumber \\
=&& 8 (1-3q+3r-u)\!\! \sum_{(\a \b \g)}\!\! \d q^{(r)}_{\a \b} \d q^{(r)}_{\b \g} \d q^{(r)}_{\g \a}  \nonumber \\
&& + 16(q-2r +u) \sum_{(\a \b)}  (\d q^{(r)}_{\a \b} )^3 ~,
\end{eqnarray}
which concludes the calculation.

%\bibliography{Lucabib.bib}
%\bibliographystyle{unsrt}

\end{document}